\documentclass[floatfix,aps,amsmath,nofootinbib,twocolumn,10pt]{revtex4}

\usepackage{listings}
\usepackage{graphicx}
\usepackage{bm}
\usepackage{rotating}
\usepackage{array}
\usepackage{amsmath}
\usepackage{amssymb} %花体字母加粗
\usepackage{mathrsfs} %花体字母
\usepackage{cancel}
\usepackage{diagbox}
\usepackage{color}

\def\({\left(}
\def\){\right)}
\def\[{\left[}
\def\]{\right]}

\def\e{\begin{equation}}
\def\q{\end{equation}}
\def\m{\begin{eqnarray}}
\def\n{\end{eqnarray}}

\begin{document}

\title{Implicit Likelihood Inference and $z$-Binned Reconstruction of Dark Energy $w(z)$} 

\author{Ke Wang$^{1}$}
\thanks{{wangke@lnnu.edu.cn}}
\author{Jia-Yi Feng$^{1}$}
\author{Jianbo Lu$^{1}$}
\affiliation{$^1$Department of Physics, Liaoning Normal University, Dalian 116029, China}

\date{\today}

\begin{abstract}
In this paper, to reconstruct the equation of state (EOS) of dark energy (DE) $w(z)$ with the redshift binning method, we first introduce a $w_i$CDM model with a piecewise-constant EOS in $7$ redshift bins.
Then, we turn to the Learning the Universe Implicit Likelihood Inference (LtU-ILI) pipeline to perform a multi-round ILI of $w_i$ from the cosmological data combination, including $TT$, $TE$, $EE$ and lensing power spectra of Planck 2018, distance ratios of DESI DR2 and corrected apparent magnitudes of SNIa from Pantheon+ sample. 
More precisely, we build the Cosmic Microwave Background (CMB) power spectrum, Baryon Acoustic Oscillation (BAO) distance ratio and Type Ia Supernovae (SNIa) apparent magnitude simulators by $\mathtt{CLASS}$ and embed them into the LtU-ILI pipeline. 
And, using Sequential Neural Likelihood Estimation (SNLE), we sequentially train neural networks with $6$ rounds of total $6\times20000$ simulations to target a ``black box'' likelihood of our forward model $w_i$CDM.
Finally, with the estimated posteriors of $w_i$, we find that except for the unconstrained $w_5$ and $w_6$ (the last two bins), our reconstruction of $w(z)$ marginally favors dynamical DE in the first bin and is consistent with the cosmological constant at $68\%$ C.L. in the other bins.
\end{abstract}

\pacs{???}

\maketitle

%%%%%%%%%%%%%%%%%%%%%%%%%%%%%%%%%%%%%%%%
%%%%%%%%%%%%%%%%%%%%%%%%%%%%%%%%%%%%%%%

\section{Introduction}
\label{sec:intro}
Dark energy (DE) was introduced to explain the accelerating expansion of the late Universe~\cite{SupernovaSearchTeam:1998fmf,SupernovaCosmologyProject:1998vns}. As the dominant component of the late Universe, DE is simply characterized by its equation of state (EOS) $w(z)$. However, for some well-known DE models, such as the $\Lambda$CDM model with a cosmological constant $w(z)=-1$, the $w$CDM model with a constant EOS $w$ and the CPL model with a $z$-dependent EOS $w(z)=w_0+w_a z/(1+z)$~\cite{Chevallier:2000qy,Linder:2002et}, their EOSs are too simplistic to avoid the Hubble tension between the local direct measurements of the Hubble constant $H_0$~\cite{Riess:2021jrx} and the Planck 2018 prediction~\cite{Planck:2018vyg}, or to capture the subtle evolutionary features during the potential dynamical evolution of $w(z)$ implied by DESI DR2~\cite{DESI:2025zgx}.
 
Therefore, it would be valuable to reconstruct the DE $w(z)$ for an accurate expansion history of the late Universe. In other words, observations of the late expansion history are also essential for reconstruction of $w(z)$. For example, one can reconstruct $w(z)$ using Gaussian processes from Type Ia Supernovae (SNIa) surveys~\cite{Seikel:2012uu}, principal component analysis from observational Hubble parameter data~\cite{Liu:2015mkm}, polynomial expansion by including Baryon Acoustic Oscillation (BAO) measurements~\cite{Gonzalez-Fuentes:2025lei,Li:2025ops}, among others. Although Cosmic Microwave Background (CMB) power spectra were formed in the early Universe where DE is expected to be negligible, taking the Planck measurement of CMB power spectra~\cite{Planck:2018vyg} into consideration allows other related background parameters to be constrained extremely accurately, hence a more accurate reconstruction of $w(z)$ indirectly.
However, to globally fit CMB power spectra during reconstruction of $w(z)$, DE extensions to the base $\Lambda$CDM model must be introduced, which can be generally denoted as $w(z)$CDM and inherently contains six base cosmological parameters.
In particular, for reconstruction of $w(z)$ with the redshift binning method in such a model, one has to deal with the large dimensionality of the parameter space. Now there are three treatments: confining to limited bins~\cite{Huang:2016fxc}; replacing CMB power spectra with CMB distance priors~\cite{Li:2025ops,Chen:2018dbv}; introducing a theoretical prior covariance between bins~\cite{Zhao:2017cud,Dai:2018zwv,Gerardi:2019obr,DESI:2025wyn}.

In this paper, we propose a novel $w(z)$ reconstruction method, which can not only perform a global fit to CMB power spectra, but also consider as many uncorrelated bins as possible.
First, we will still use the $z$-binned reconstruction of $w(z)$, where $w(z)$ is parameterized as $w_i$ in each redshift bin. Then, this $w_i$CDM model can be used to globally fit CMB power spectra. However, since traditional explicit likelihood inference scales poorly with the dimensionality of the parameter space, a large number of $w_i$ with six base cosmological parameters cannot be estimated together.
Here we turn to a novel alternative, Implicit Likelihood Inference (ILI)~\cite{Cranmer:2019eaq}, which is also known as Likelihood Free Inference (LFI) and Simulation Based Inference (SBI). 
Compared to traditional explicit likelihood inference, instead of sampling the full posterior of the joint first and marginalizing it then, ILI can directly learn 1- and 2-dimensional marginal posteriors.
More precisely, given an initial wider cosmological parameter priors, the efficient CMB simulator $\mathtt{CLASS}$~\cite{Blas:2011rf} or $\mathtt{CAMB}$~\cite{Lewis:1999bs} can forward-simulate sets of parameter values to the training data of $TT$, $TE$, $EE$ and CMB lensing-potential power spectra.
After the first round of training or learning the statistical relationship between parameters and data by machine learning (ML), the obtained or learned posteriors serve as the new priors for the second round of training, and so on. This sequential learning has been applied to ILI of the cosmological parameters from CMB data~\cite{Cole:2021gwr}.

This paper is organized as follows. In Section~\ref{sec:simulator}, we introduce the CMB power spectrum, BAO distance ratio and SNIa apparent magnitude simulators for $w_i$CDM model. In Section~\ref{sec:ili}, we perform a multi-round ILI of $w_i$ and six base cosmological parameters. Finally, a brief summary and discussion are provided in Section~\ref{sec:sum}.

\section{CMB, BAO and SNIa simulators}
\label{sec:simulator}
Given a specific DE extension to $\Lambda$CDM with parameters $\bm{\theta}$, the corresponding CMB, BAO or SNIa simulator can forward-simulate candidate parameter values drawn from the prior $\mathcal{P}(\bm{\theta})$ to observations.
For reconstruction of DE's EOS $w(z)$ with the redshift binning method, we choose the DE extension to $\Lambda$CDM as $w_i$CDM model, whose DE EOS is
\begin{equation}
w(z)=w_i,~~~z_i<z\le z_{i+1},~~~i=0,1,2...\,,
\end{equation}
where a piecewise-constant parameterization is adopted.
This parameterization provides a flexible and model-independent approach for reconstructing the dynamical evolution of DE.
To check the evolutionary features of $w(z)$ suggested by DESI DR2~\cite{DESI:2025zgx}, the redshift bins are chosen as
\begin{equation}
z_i=\{0,\,0.4,\,0.6,\,0.8,\,1.1,\,1.4,\,1.9,\,\infty\},
\end{equation}
so that each bin contains one distance ratio from DESI DR2~\cite{DESI:2025zgx}.
For such a piecewise-constant EOS, the DE density evolves as 
\begin{equation}
\rho_{\rm DE}(z)=\rho_{\rm DE}(z_i)\left(\frac{1+z}{1+z_i}\right)^{3(1+w_i)},~~~z_i<z\le z_{i+1},
\end{equation}
where $\rho_{\rm DE}(z_i)$ is the DE density at the redshift $z=z_i$ and $w_i$ is a constant EOS defined in bin $z_i<z\le z_{i+1}$.
While $w(z)=w_i$ is allowed to vary discontinuously between bins, $\rho_{\rm DE}(z)$ evolves recursively across neighboring bins and remains continuous at each bin boundary. Finally, DE contributes to the cosmic expansion through the Friedmann equation,
\begin{equation}
\frac{H^2(z)}{H_0^2}=\Omega_{\rm r} (1+z)^{4}
+\Omega_{\rm m} (1+z)^{3}
%+\Omega_k (1+z)^{2}
+\Omega_{\rm DE}
\frac{\rho_{\rm DE}(z)}
{\rho_{\rm DE}(0)},
\end{equation}
where $\Omega_{\rm r}$, $\Omega_{\rm m}$ and $\Omega_{\rm DE}$ denote today fractional energy densities relative to critical in radiation, matter and DE, respectively.

Since BAO and SNIa measurements constrain the late expansion history well, besides CMB power spectrum simulator, we will also create BAO distance ratio and SNIa apparent magnitude simulators.
As a result, there would be 14 parameters: $6$ base cosmological parameters and $7$ $w_i$ of $w_i$CDM for all three simulators and an additional nuisance parameter, the absolute magnitude of SNIa $M$, only for the SNIa simulator.
For the sequential learning approach, their initial uniform priors can be wider
\begin{equation}
\label{eq:prior} 
\bm{\theta}=\left\{
\begin{aligned}
&\omega_{\rm b}\\
&\omega_{\rm c}\\
&H_0\rm{[km~s^{-1}~Mpc^{-1}]} \\
&\ln (10^{10}A_{\rm s})\\
&n_{\rm s}\\
&\tau\\
&w_0\\
&w_1\\
&w_2\\
&w_3\\
&w_4\\
&w_5\\
&w_6\\
&M
\end{aligned}
\right\}= 
\left\{
\begin{aligned}
&\mathcal{U}(0.020,0.024)\\
&\mathcal{U}(0.11,0.13)\\
&\mathcal{U}(50,100)\\
&\mathcal{U}(2.98,3.20)\\
&\mathcal{U}(0.94,1.02)\\
&\mathcal{U}(0.04,1.00)\\
&\mathcal{U}(-3,0)\\
&\mathcal{U}(-3,3)\\
&\mathcal{U}(-10,3)\\
&\mathcal{U}(-10,3)\\
&\mathcal{U}(-10,3)\\
&\mathcal{U}(-50,3)\\
&\mathcal{U}(-100,0)\\
&\mathcal{U}(-20,-18)
\end{aligned}
\right\}.
\end{equation}
For the first $6$ base cosmological parameters, their priors are wider than $\pm5\sigma$ ranges of constraints from the combination of Planck 2018 and BAO~\cite{Planck:2018vyg} by traditional explicit likelihood inference; for $w_i$ at higher redshift bins, their priors are much wider because of the sparse BAO and SNIa data there.

\subsection{CMB power spectrum simulator}
As proposed in~\cite{Cole:2021gwr}, the CMB power spectrum simulator should consist of the calculation of CMB power spectra $C_{\ell}^{XX'}(\bm{\theta})$ ($X=T,E$) with $w_i$CDM model and the corresponding noise realizations.
The former can be done using the Einstein-Boltzmann solver $\mathtt{CLASS}$~\cite{Blas:2011rf} given a set of candidate parameter values from the prior $\mathcal{P}(\bm{\theta})$.

As for the noise realizations, two contributions are involved: instrumental noise and cosmic variance.
1) For the Planck experiment, its noise spectrum can be approximated by the combination of a Gaussian beam and a Gaussian white noise~\cite{CORE:2016npo}
\begin{equation}
N_{\ell}^{XX'}=\delta_{XX'}\theta_{\rm FWHM}^2\sigma_X^2\exp\left(\ell(\ell+1)+\frac{\theta_{\rm FWHM}^2}{8\ln2}\right),
\end{equation}
where $\sigma_X$ is the detector sensitivity and $\theta_{\rm FWHM}$ is the full width at half maximum of a Gaussian beam. 
For simplicity, we just turn to a ready-made $N_{\ell}^{XX'}$ for Planck's temperature and polarization measurements, $\mathtt{fake\_planck\_realistic}$ in $\mathtt{MontePython}$~\cite{Audren:2012wb}.
Now, we have $\bar{C}_{\ell}^{XX'}(\bm{\theta})=C_{\ell}^{XX'}(\bm{\theta})+N_{\ell}^{XX'}$. 

\begin{figure*}[]
\begin{center}
\includegraphics[width= 8.5cm]{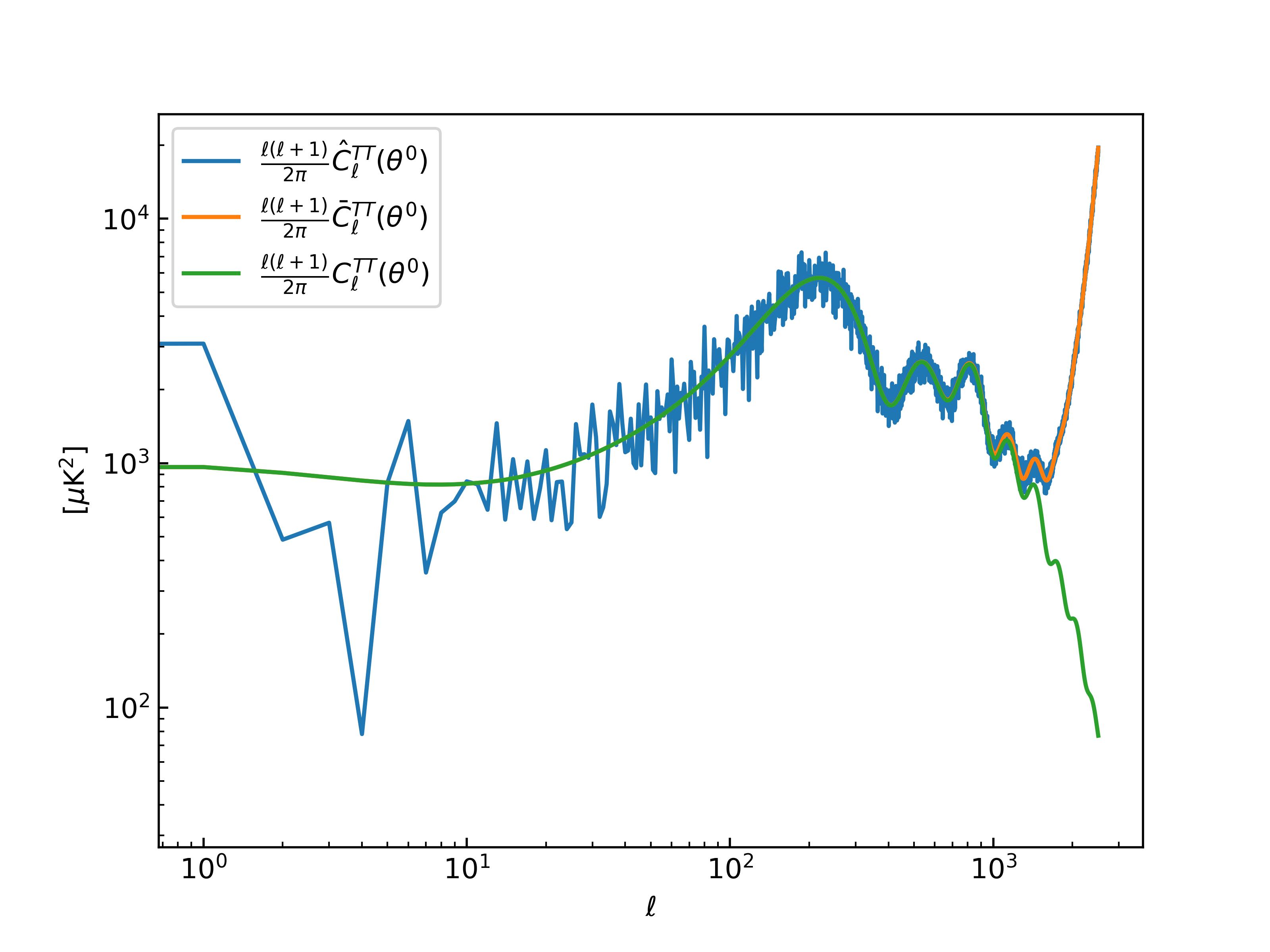}
\includegraphics[width= 8.5cm]{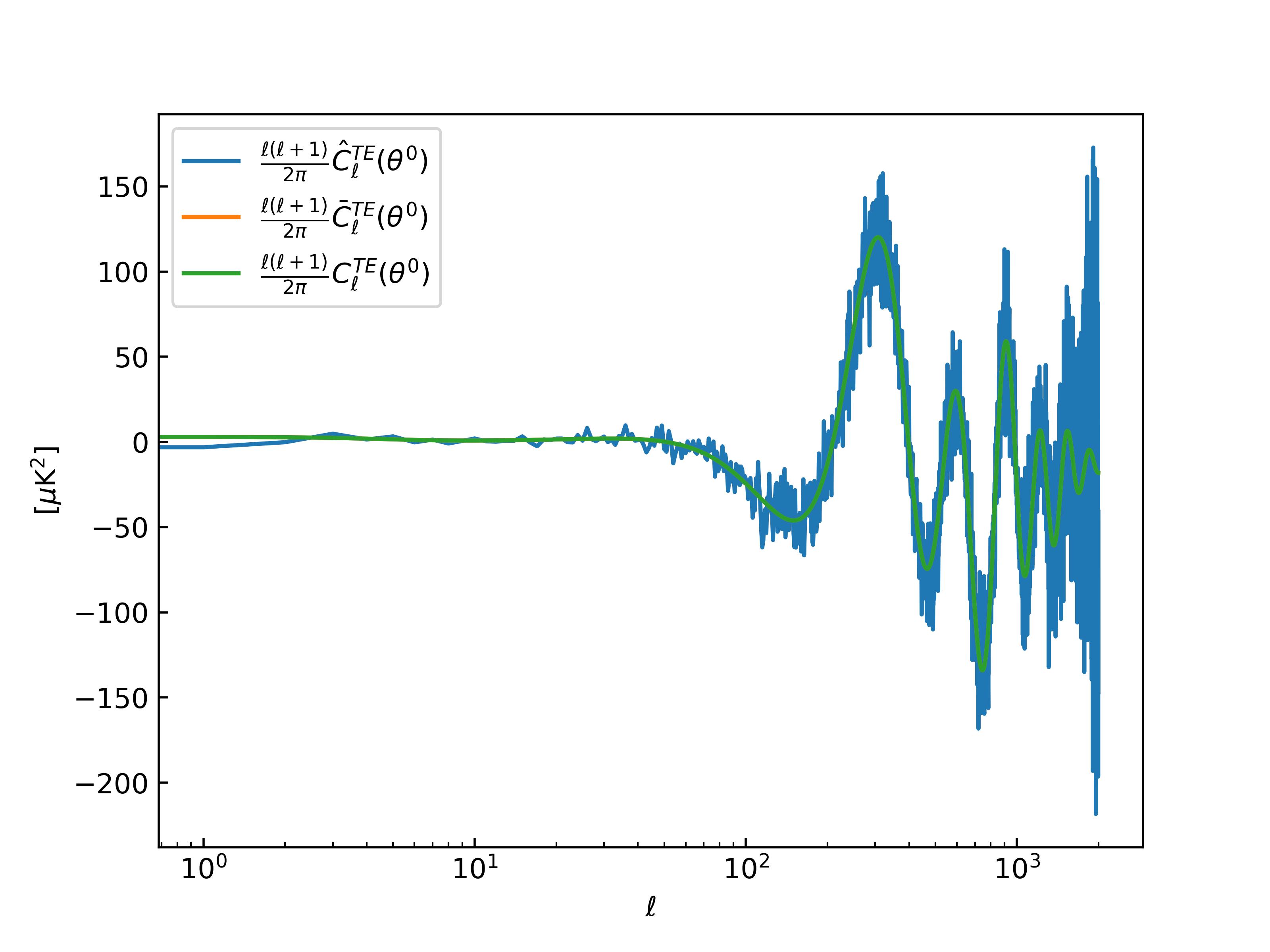}
\includegraphics[width= 8.5cm]{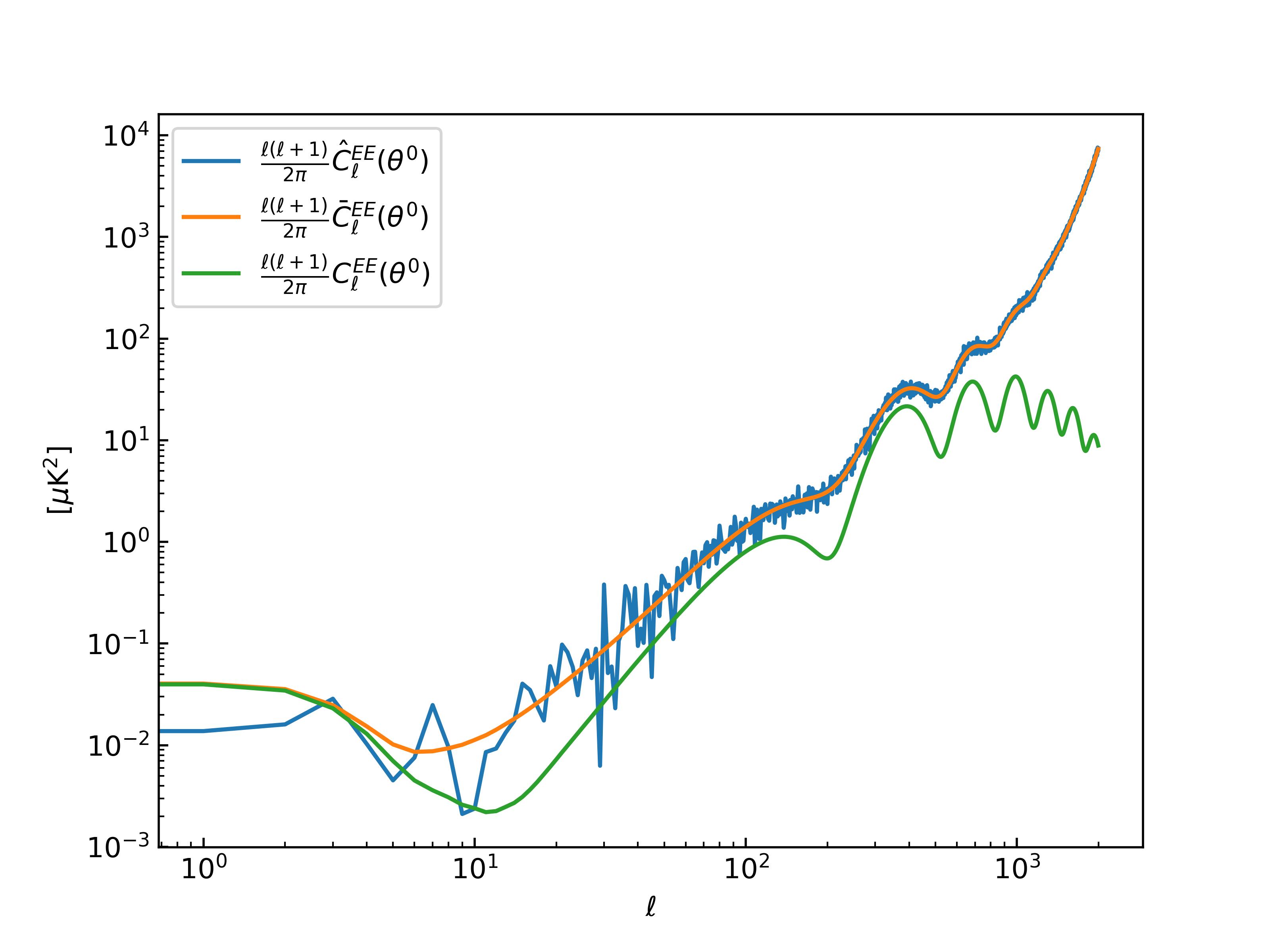}
\end{center}
\caption{Comparison among $C_{\ell}^{XX'}(\bm{\theta}^0)$, $\bar{C}_{\ell}^{XX'}(\bm{\theta}^0)$ and $\hat{C}_{\ell}^{XX'}(\bm{\theta}^0)$, respectively. Since $\bm{\theta}^0$ only contains six base cosmological parameters of the fiducial cosmology, plots are almost the same as those in~\cite{Wang:2025xjc}.}
\label{fig:Cls}
\end{figure*}

2) According to the estimator of power spectra
\begin{equation}
\hat{C}_{\ell}^{XX'}(\bm{\theta})=\frac{1}{2\ell+1}\sum_{m=-\ell}^{\ell}a_{\ell m}^{X*}a_{\ell m}^{X'},
\end{equation} 
there obviously is a fundamental uncertainty (due to cosmic variance) in knowledge about the power spectra.
In principle, the multipole coefficients $a_{\ell m}^{X}$ of CMB maps are assumed to follow a Gaussian distribution and can be sampled according to the covariance matrix 
\begin{equation}
\label{eq:covmatrix}
{\rm Cov}_{a_{\ell m}^{X}}=\left(
\begin{aligned}
&\bar{C}_{\ell}^{TT}(\bm{\theta})~~\bar{C}_{\ell}^{TE}(\bm{\theta})\\
&\bar{C}_{\ell}^{TE}(\bm{\theta})~~\bar{C}_{\ell}^{EE}(\bm{\theta})
\end{aligned}
\right).
\end{equation}
Especially, for a specific couple of $\ell'$ and $m'$, when its block in ${\rm Cov}_{a_{\ell m}^{X}}$ is decomposed into the product of a $2\times2$ lower triangular matrix $L_2$ and its conjugate transpose $L_2^T$ using the Cholesky decomposition, $a_{\ell' m'}^{X}$ can be simply sampled as 
\begin{equation}
\left(
\begin{aligned}
&a_{\ell' m'}^{T}\\
&a_{\ell' m'}^{E}
\end{aligned}
\right)=L_2(\bm{\theta}) 
\left(
\begin{aligned}
&\mathcal{N}(0,1)\\
&\mathcal{N}(0,1)
\end{aligned}
\right).
\end{equation}
In practice, however, we treat $\hat{C}_{\ell}^{XX'}(\bm{\theta})$ directly, instead of sampling $a_{\ell m}^{X}$.
More precisely, we directly sample $\hat{C}_{\ell}^{XX'}(\bm{\theta})$ from a Wishart distribution~\cite{Percival:2006ss} at low $\ell$ and set $\hat{C}_{\ell}^{XX'}(\bm{\theta})=\bar{C}_{\ell}^{XX'}(\bm{\theta})+n_{\ell}^{XX'}(\bm{\theta}^0)$ at $\ell>52$ because a Wishart distribution tends towards a multivariate Gaussian distribution at large $\ell$. 
Here $n_{\ell}^{XX'}(\bm{\theta}^0)$ depends on a fiducial cosmology\footnote{Our noise realizations of Planck 2018, $n_{\ell}^{XX'}(\bm{\theta}^0)$, are model-independent. Given Planck 2018 data, the theoretical power spectra predicted by different best-fit models must be similar. Therefore, the choice of fiducial model does not matter only if its theoretical power spectra fit Planck 2018 data well. Here the fiducial cosmology is the $\Lambda$CDM model with $\bm{\theta}^0=\{\omega_{\rm b},\omega_{\rm c},100\theta_s,\ln (10^{10}A_{\rm s}),n_{\rm s},\tau\}=\{0.02237,0.1200,1.04092,3.044,0.9649,0.0544\}$. Of course, one can also directly download its theoretical power spectra $C_{\ell}^{XX'}(\bm{\theta}^0)$ from Planck 2018~\cite{PLA}.} and can be simply sampled from a multivariate normal distribution with the covariance matrix
\begin{align}
\nonumber
&{\rm Cov}_{n_{\ell}^{XX'}}=\frac{2}{2\ell+1}\times\\
&\left(
\begin{aligned}
&\left(\bar{C}_{\ell}^{TT}(\bm{\theta}^0)\right)^2~~~~~~~~~~\bar{C}_{\ell}^{TT}(\bm{\theta}^0)\bar{C}_{\ell}^{TE}(\bm{\theta}^0)~~~~~~~~~\left(\bar{C}_{\ell}^{TE}(\bm{\theta}^0)\right)^2\\
&~~~~~~~~~~~~~~~~~~~~~~~~~~~~~~~~\left.\frac{1}{2}\right(\\
&\bar{C}_{\ell}^{TT}(\bm{\theta}^0)\bar{C}_{\ell}^{TE}(\bm{\theta}^0)~~\bar{C}_{\ell}^{TT}(\bm{\theta}^0)\bar{C}_{\ell}^{EE}(\bm{\theta}^0)~~\bar{C}_{\ell}^{TE}(\bm{\theta}^0)\bar{C}_{\ell}^{EE}(\bm{\theta}^0)\\
&~~~~~~~~~~~~~~~~~~~~~~~~~+\left.\left(\bar{C}_{\ell}^{TE}(\bm{\theta}^0)\right)^2\right)\\
&\left(\bar{C}_{\ell}^{TE}(\bm{\theta}^0)\right)^2~~~~~~~~~~\bar{C}_{\ell}^{TE}(\bm{\theta}^0)\bar{C}_{\ell}^{EE}(\bm{\theta}^0)~~~~~~~~~\left(\bar{C}_{\ell}^{EE}(\bm{\theta}^0)\right)^2
\end{aligned}
\right).
\end{align}
In Fig.~\ref{fig:Cls}, we plot $C_{\ell}^{XX'}(\bm{\theta}^0)$, $\bar{C}_{\ell}^{XX'}(\bm{\theta}^0)$ and $\hat{C}_{\ell}^{XX'}(\bm{\theta}^0)$, respectively.

Our CMB simulator also includes the CMB lensing simulation. Besides $w_i$CDM's CMB lensing prediction by $\mathtt{CLASS}$~\cite{Blas:2011rf}, the corresponding noise realizations of the lensing power spectrum at $9$ weighted multipole-bin centres are sampled from a multivariate normal distribution, which is derived from the $9\times9$ covariance matrix $\mathtt{smi\dots\_ndclpp\_p\_teb\_consext8\_cov}$ of Planck 2018 minimum-variance (MV) lensing band powers at multipole range $8\sim400$~\cite{Planck:2018lbu,PLA}.

\subsection{BAO distance ratio simulator}
BAO measurements can be summarized by distance ratios, such as the transverse comoving
distance to the Hubble distance $D_{\rm M}/D_{\rm H}$, which are defined as
\begin{align}
D_{\rm M}(z)&=\frac{c}{H_0}\int_0^z\frac{dz'}{H(z')/H_0}~~({\rm flat~Universe}),\\
D_{\rm H}(z)&=\frac{c}{H(z)},
\end{align}
and the isotropic BAO distance to the scale of the pre-recombination sound horizon $D_{\rm V}/r_{\rm d}$, which are defined as
\begin{align}
D_{\rm V}(z)&\equiv\left(cD_{\rm M}(z)^2D_{\rm H}(z)\right)^{1/3},\\
r_{\rm d}&=\int_{z_{\rm d}}^{\infty}\frac{c_{\rm s}(z)}{H(z)}dz,
\end{align}
where $c_{\rm s}(z)$ is the speed of sound in the photon-baryon fluid and $z_{\rm d}$ is the redshift when the photons stop dragging the baryons.
Since DESI DR2~\cite{DESI:2025zgx} provides these distance ratios and their covariance matrices at redshifts $z=\{0.295,0.510,0.706,0.934,1.321,1.484,2.330\}$, $w_i$CDM will also theoretically calculate these distance ratios with $\mathtt{CLASS}$~\cite{Blas:2011rf} at the same redshifts.
The BAO distance ratio simulator just combines the theoretical predictions of these distance ratios by $w_i$CDM and the corresponding noise realizations that are sampled from the multivariate normal distribution of DESI DR2 covariance matrix.

\subsection{SNIa apparent magnitude simulator}
One of SNIa measurements is the apparent magnitudes of SNIa at specific redshifts
\begin{equation}
m(z)=5\log_{10}\left[\frac{D_{\rm L}(z)}{1{\rm Mpc}}\right]+25+M,
\end{equation}
where $M$ is the absolute magnitude of SNIa and $D_{\rm L}$ is the luminosity distance
\begin{equation}
D_{\rm L}(z)=(1+z)D_{\rm M}(z).
\end{equation}
Pantheon+ sample~\cite{Brout:2022vxf} derives the corrected $m$ from $1701$ light curves of $1550$ distinct SNIa in the redshift range $0.001<z<2.26$ and also provides a $1701\times1701$ covariance matrix.
To efficiently sample the noise realizations in the SNIa apparent magnitudes simulator, we first perform the Cholesky decomposition of this $1701\times1701$ covariance matrix into a $1701\times1701$ lower triangular matrix $L_{1701}$ and its conjugate transpose $L_{1701}^T$.
Then the noise realizations can be efficiently sampled from
\begin{equation}
\left(
\begin{aligned}
&n_1\\
&~~\vdots\\
&n_{1701}
\end{aligned}
\right)=
L_{1701} 
\left(
\begin{aligned}
&\mathcal{N}_1(0,1)\\
&~~~\vdots\\
&\mathcal{N}_{1701}(0,1)
\end{aligned}
\right).
\end{equation}
Finally, the SNIa apparent magnitude simulator combines $w_i$CDM's theoretical predictions of $m(z)$ at the same $1701$ redshifts by $\mathtt{CLASS}$~\cite{Blas:2011rf} and their corresponding noise realizations.

\section{Multi-round ILI and reconstruction of $w_i$}
\label{sec:ili}  
In this paper, the Learning the Universe Implicit Likelihood Inference ($\mathtt{LtU}$-$\mathtt{ILI}$) pipeline~\cite{Ho:2024whi} is used: a $\mathtt{PyTorch}$~\cite{PyTorch} package, $\mathtt{sbi}$~\cite{sbi}, as the computational backend and the sequential analog of Neural Likelihood Estimation (SNLE)~\cite{Alsing:2018eau,Papamakarios:2019ccu} as the inference engine.
For the NLE model, its loss function is 
\begin{equation}
\mathcal{L}=-\mathbb{E}_{\mathcal{D_{\rm train}}}[\ln q_{\bm W}({\bm x}|{\bm \theta})],
\end{equation}
where $\mathbb{E}_{\mathcal{D_{\rm train}}}$ is an expectation taken over all data-parameter pairs $\{({\bm x}_i,{\bm \theta}_i)\}$ of the training data set $\mathcal{D_{\rm train}}$, ${\bm W}$ is the network weights and $q_{\bm W}({\bm x}|{\bm \theta})$ converges to the true likelihood by minimizing of $\mathcal{L}$ over ${\bm W}$. 
Then, we selected the following training details based on its best validation performance: 
the NLE model consists of an ensemble of $6$ Neural Density Estimations~\cite{Papamakarios:2019ccu} (NDEs); each NDE uses a Masked Autoregressive Flow~\cite{MAF} (MAF) architecture with $100$ hidden features and $5$ transformations;
in each training round of the NLE model, the training data set $\mathcal{D_{\rm train}}$ contains $20000$ simulated data-parameter pairs $\{({\bm x}_i,{\bm \theta}_i)\}$;
total $6$ rounds of sequential training are performed.

After the last training round, $q_{\bm W}({\bm x}|{\bm \theta})$ serves as the learned likelihood and consequently an amortized posterior $\mathcal{P}({\bm \theta}|{\bm x})\propto q_{\bm W}({\bm x}|{\bm \theta})\mathcal{P}({\bm \theta})$ is learned. 
Here, the local amortization of $\mathcal{P}({\bm \theta}|{\bm x})$ means that the marginal posterior of $\bm\theta$ can be sampled from $\mathcal{P}({\bm \theta}|{\bm x})$ without retraining, for $\bm x$ not from $\mathcal{D_{\rm train}}$.
For example, for $\{({\bm x}_i,{\bm \theta}_i)\}$ from another different testing data set $\mathcal{D_{\rm test}}$, their marginal posteriors can also be sampled directly from $\mathcal{P}({\bm \theta}|{\bm x})$ according to local amortization.
Therefore, some special validation tests can be done, such as the rank statistics, the marginal percentile coverage test and the direct comparison between true and predicted parameter values. These tests are not accessible to traditional explicit likelihood inference.

\begin{figure*}[]
\begin{center}
\includegraphics[width= 18cm]{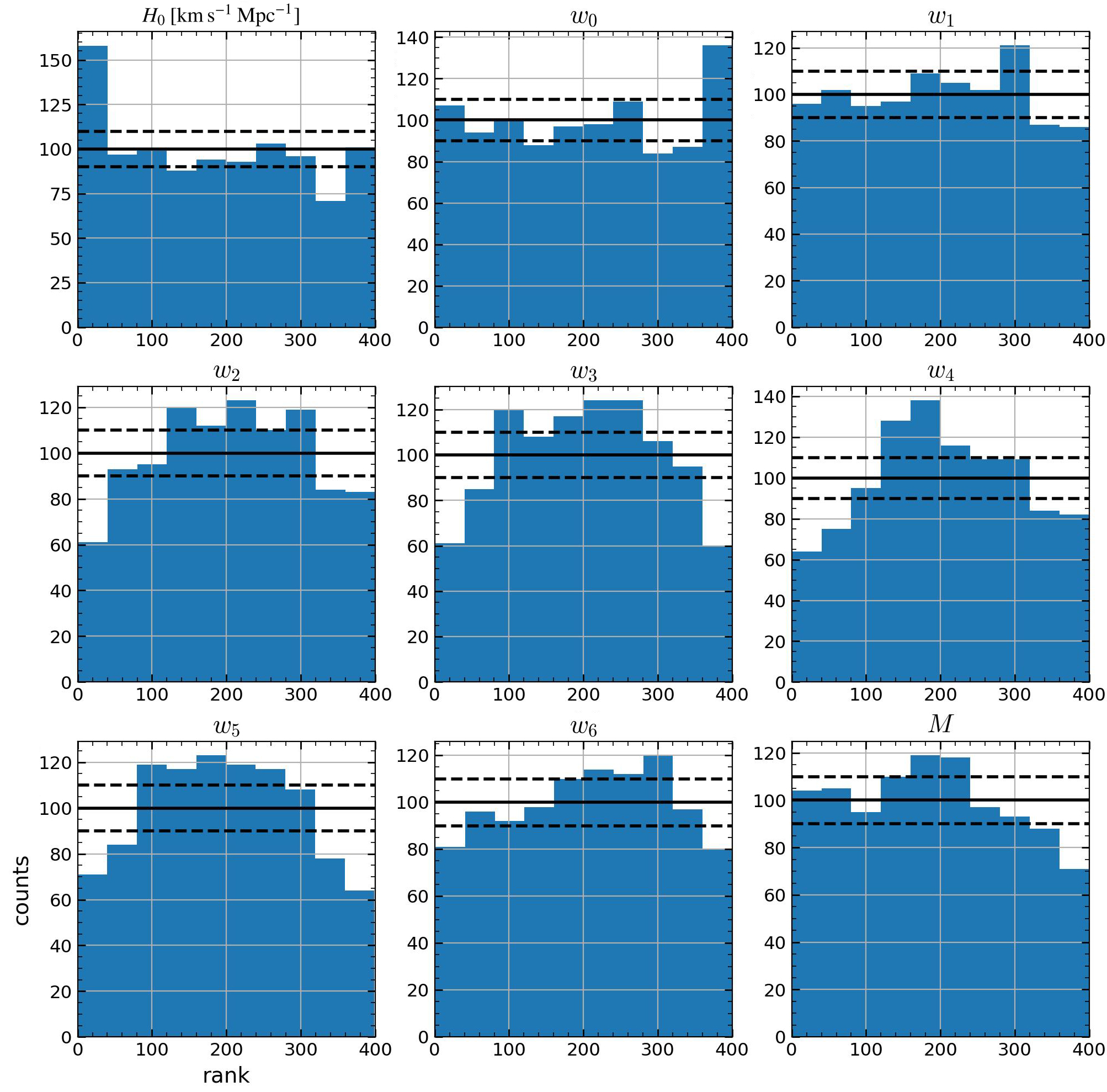}
\end{center}
\caption{Rank statistic of $w_i$, $H_0$ and $M$, where the rank range $1\text{--}400$ results from the comparison between ${\theta}_i$ and its $N=400$ posterior samples ${\theta}'_i$ (Eq.~\ref{eq:rank}) and $\sum {\rm counts} = 1000$ is equal to the total number of data-parameter pairs in the testing data set $\mathcal{D_{\rm test}}$.
The distribution of rank statistic and its deviation from $\mathcal{U}(0,1)$ can diagnose the incorrectness of estimated posterior~\cite{Wang:2025xjc,Hahn:2022wgo,Talts:2018zdk,Hahn:2022nda}: 
when $\mathcal{P}({\bm \theta}|{\bm x})$ is exactly estimated, the rank statistic distributes uniformly (black solid);
while a $\cap$-shape rank statistic suggests that the estimated $\mathcal{P}({\bm \theta}|{\bm x})$ is broader than the true posterior, an $\cup$-shaped rank statistic implies a narrower estimated $\mathcal{P}({\bm \theta}|{\bm x})$;
the estimated $\mathcal{P}({\bm \theta}|{\bm x})$ is biased in the opposite direction with respect to the true posterior when an asymmetric rank histogram appears;
obvious (anti-)correlations between parameters lead to the spikes at the boundaries of histogram.}
\label{fig:rank}
\end{figure*}

\begin{figure*}[]
\begin{center}
\includegraphics[width= 18cm]{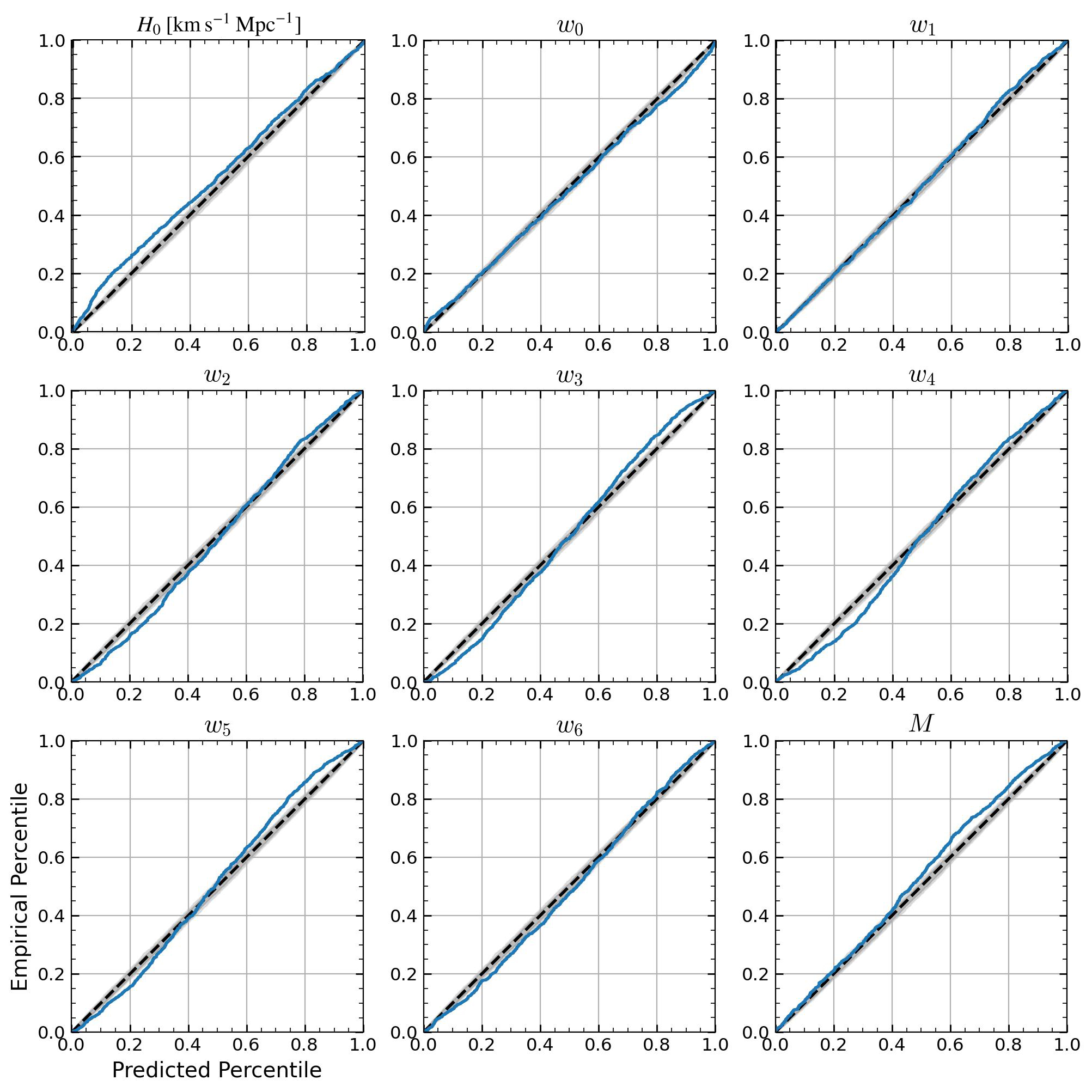}
\end{center}
\caption{Marginal percentile coverage test for $w_i$, where the predicted percentile results from the distribution of all CDF values calculated with $({\bm x}_i,{\bm \theta}_i)\in\mathcal{D_{\rm test}}$ in Eq.~(\ref{eq:CDF}) and the empirical percentile represents uniform distribution $\mathcal{U}(0,1)$.}
\label{fig:pp}
\end{figure*}

\begin{figure*}[]
\begin{center}
\includegraphics[width= 18cm]{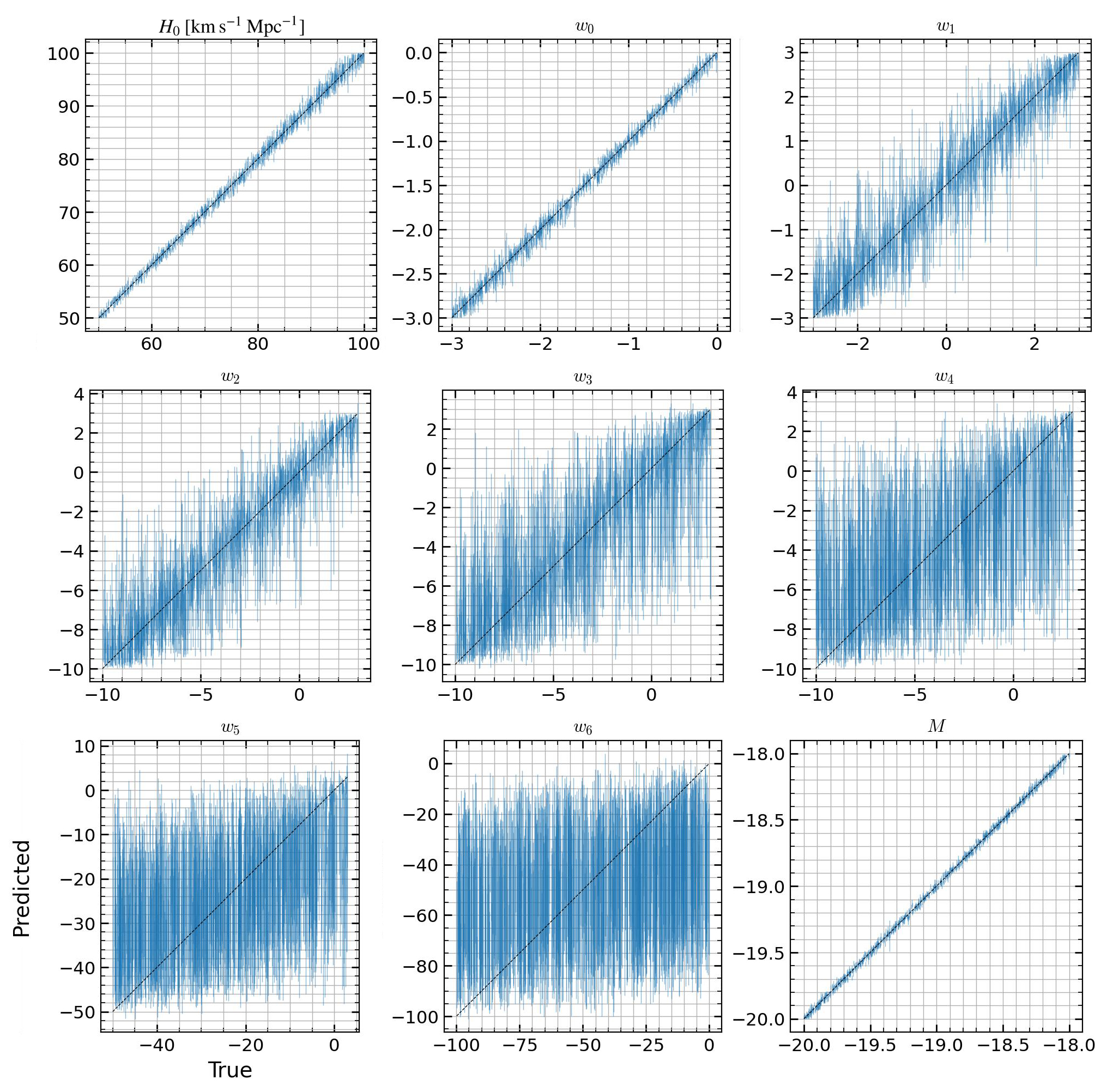}
\end{center}
\caption{Comparison between predicted and true values of $w_i$.}
\label{fig:TvsP}
\end{figure*}

For $\mathcal{D_{\rm test}}$ including $1000$ data-parameter pairs $({\bm x}_i,{\bm \theta}_i)$, we can count any $\theta_i$'s rank in its $N=400$ posterior samples
\begin{equation}
\label{eq:rank}
\{\theta'_1,\theta'_2,...,\theta'_{N=400}\}\sim\mathcal{P}({\theta}|{\bm x}_i),~\forall ({\bm x}_i,{\bm \theta}_i)\in\mathcal{D_{\rm test}},
\end{equation}
where $\theta'_i$ are ordered in ascending numerical order. For example, $\theta_i$ is ranked third when $\theta'_2<\theta_i<\theta'_3$.
The rank statistics is based on the conclusion that if the true posterior is exactly estimated, the rank statistic of $\theta_i$ should be distributed uniformly~\cite{Wang:2025xjc,Hahn:2022wgo,Talts:2018zdk,Hahn:2022nda}. 
As shown in Fig.~\ref{fig:rank}, there is an obvious left or right spike for $H_0$ or $w_0$, which results from the anti-correlation
between them. Except for spikes, the individual rank statistic of $H_0$, $w_0$ or $w_1$ is distributed uniformly. The $\cap$-shape rank statistics of $w_2\text{--}w_6$ mean that their estimated posteriors are broader than the true ones.
Changing the rank with the cumulative density function (CDF) value, 
\begin{equation}
\label{eq:CDF}
\int_{-\infty}^{\theta_i}d\theta\mathcal{P}({\theta}|{\bm x}_i)\sim\mathcal{U}(0,1),~\forall ({\bm x}_i,{\bm \theta}_i)\in\mathcal{D_{\rm test}},
\end{equation}
there would be a distribution of all CDF values calculated with $({\bm x}_i,{\bm \theta}_i)\in\mathcal{D_{\rm test}}$.
Similarly, this distribution would also be distributed uniformly when the true posterior is exactly estimated.
So, we can compare this distribution with $\mathcal{U}(0,1)$ using their own CDF, as shown in the percentile-percentile (P-P) plot (Fig.~\ref{fig:pp}). The disagreement for $H_0$ or $w_i$ results from the spike in their own rank statistic. The good agreement for $w_1$ means that its learned posterior is almost consistent with the truth. Also, the broader estimated posteriors of $w_2\text{--}w_6$ suggested by rank statistics lead to the corresponding disagreement in the P-P plot.
Finally, in Fig.~\ref{fig:TvsP}, we directly compare the predicted values sampled from $\mathcal{P}({\theta}|{\bm x}_i)$ and the corresponding true value of $\theta_i$ in $\mathcal{D_{\rm test}}$. 
Obviously, as $z$ increases, $w_i$'s estimated posterior becomes broader, especially $w_4\text{--}w_6$.

\begin{figure*}[]
\begin{center}
\includegraphics[width= 19cm]{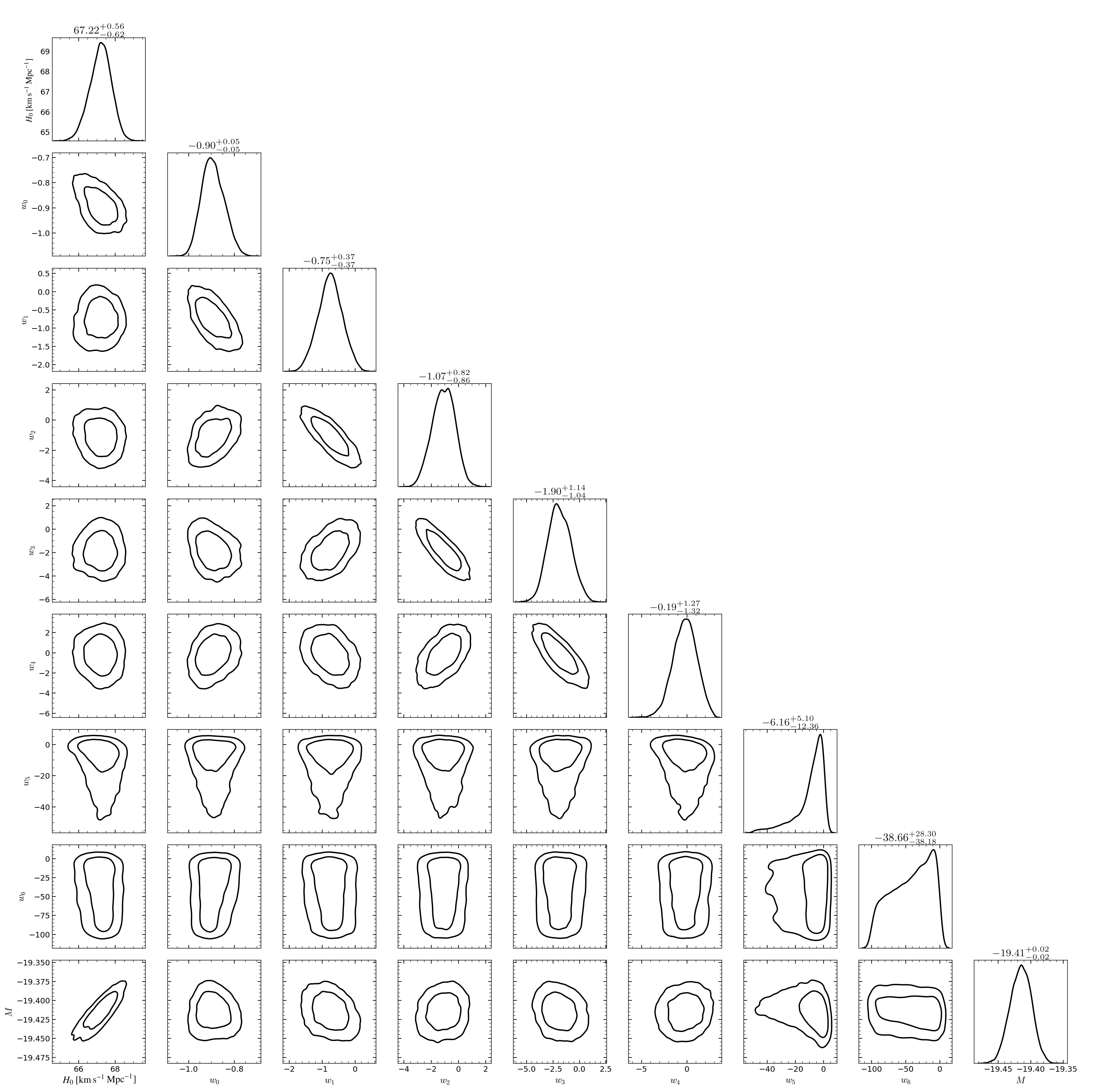}
\end{center}
\caption{Constraints on parameters of $w_i$ from Planck 2018~\cite{Planck:2018vyg}, DESI DR2~\cite{DESI:2025zgx} and Pantheon+ sample~\cite{Brout:2022vxf}, where $68\%$ interval of each parameter is displayed as the titles of diagonal subplots and contours contain $68\%$ and $95\%$ of the probability.}
\label{fig:posterior}
\end{figure*}

\begin{figure*}[]
\begin{center}
\includegraphics[width= 18cm]{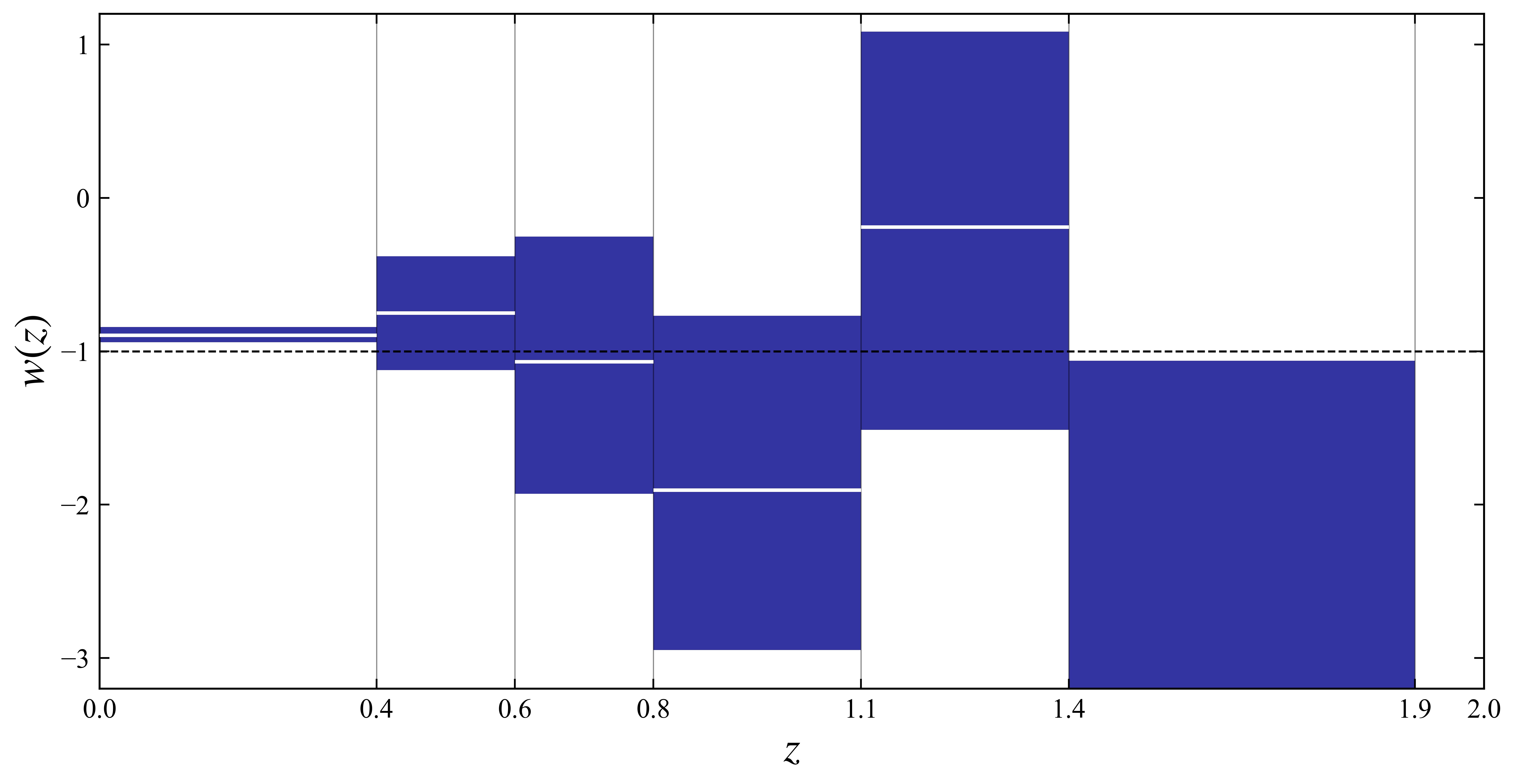}
\end{center}
\caption{Reconstruction of $w(z)$ with $w_i$, where $w_5$ and $w_6$ were not constrained well so that they deviate from $-1$ a lot and don't appear in plot, the shaded regions are $68\%$ C.L. and the white lines indicate the best-fit reconstruction of $w(z)$.}
\label{fig:wz}
\end{figure*}

Unlike $\mathcal{D_{\rm test}}$ with $1000$ data-parameter pairs $({\bm x}_i,{\bm \theta}_i)$, we only have one observed cosmological data $\bm{x}_o$, including CMB $TT$, $EE$ and $TE$ spectra and their uncertainties $\mathtt{COM\_PowerSpect\_CMB\_XX'}$ from Planck 2018~\cite{PLA}, CMB minimum-variance lensing band powers and their uncertainties $\mathtt{smi\dots\_ndclpp\_p\_teb\_consext8\_bandpowers}$ from Planck 2018~\cite{PLA}, the distance ratios and their covariance matrix from DESI DR2~\cite{DESI:2025zgx} and the corrected apparent magnitudes of SNIa and their covariance matrix from Pantheon+ sample~\cite{Brout:2022vxf}.
Due to the local amortization of learned $\mathcal{P}({\bm \theta}|{\bm x})$, the posterior of all parameters of $w_i$CDM for $\bm{x}_o$ can be sampled by the usual Monte Carlo Markov Chain sampling. Here we sample them with $\mathtt{emcee}$~\cite{Foreman-Mackey:2012any}. And we list the constraints on them at $68\%$ C.L. in Tab.~\ref{tab:results}. Also, in Fig.~\ref{fig:posterior}, we show the posterior distribution of $w_i$. We find that $w_5$ and $w_6$ cannot be constrained by data and there are obvious anti-correlations among $w_0\text{--}w_4$, between $H_0$ and $w_0$ and correlations between $H_0$ and $M$.
For clarity, we show the reconstruction of $w(z)$ with $w_i$ in Fig.~\ref{fig:wz}. Except for the unconstrained $w_5$ and $w_6$, our reconstruction of $w(z)$ marginally favors dynamical DE in the first bin and is consistent with the cosmological constant at $68\%$ C.L. in the other bins, even though the best-fit $w(z)$ profile (white line) hints at dynamical DE.

\begin{table}[t]
\centering
\caption{Constraints on parameters of $w_i$CDM from Planck 2018~\cite{Planck:2018vyg}, DESI DR2~\cite{DESI:2025zgx} and Pantheon+ sample~\cite{Brout:2022vxf}.}
\footnotesize
\renewcommand{\arraystretch}{1.3}
\setlength{\tabcolsep}{4pt}
\begin{tabular}{@{}lclc@{}}
\hline\hline
Parameter & $68\%$ C.L. & Parameter & $68\%$ C.L. \\
\hline
$\omega_{\rm b}$
&$0.02221^{+0.00014}_{-0.00014}$
&$w_0$
&$-0.90^{+0.05}_{-0.05}$
\\
$\omega_{\rm c}$
&$0.1221^{+0.0012}_{-0.0013}$
&$w_1$
&$-0.75^{+0.37}_{-0.37}$
\\
$H_0\rm{[km~s^{-1}~Mpc^{-1}]}$
&$67.22^{+0.56}_{-0.62}$
&$w_2$
&$-1.07^{+0.82}_{-0.86}$
\\
$\ln (10^{10}A_{\rm s})$
&$3.059^{+0.007}_{-0.007}$
&$w_3$
&$-1.90^{+1.14}_{-1.04}$
\\
$n_{\rm s}$
&$0.9578^{+0.0035}_{-0.0034}$
&$w_4$
&$-0.19^{+1.27}_{-1.32}$
\\
$\tau$
&$0.0581^{+0.0042}_{-0.0036}$
&$w_5$
&$-6.16^{+5.10}_{-12.36}$
\\
$M$
&$-19.41^{+0.02}_{-0.02}$
&$w_6$
&$-38.66^{+28.30}_{-38.18}$
\\
\hline
\label{tab:results}
\end{tabular}
\end{table}

\section{Summary and discussion}
\label{sec:sum} 
In this paper, we first construct the CMB power spectrum, BAO distance ratio and SNIa apparent magnitude simulators by decorating the $\mathtt{CLASS}$ theoretical outputs in $w_i$CDM model with the corresponding noise realizations for each cosmological measurements.
Then, we learn an amortized posterior $\mathcal{P}({\bm \theta}|{\bm x})$ by the inference engine of SNLE after $6$ rounds of training neural networks with total $6\times20000$ data-parameter pairs $({\bm x}_i,{\bm \theta}_i)$.
Moreover, we also perform the validation test with $\mathcal{D_{\rm test}}$, such as the rank statistics, the P-P plot and the percentile coverage plot.
With the credible learned $\mathcal{P}({\bm \theta}|{\bm x})$, $\mathcal{P}({\bm \theta}|{\bm x}_0)$ is sampled with $\mathtt{emcee}$~\cite{Foreman-Mackey:2012any}, where ${\bm x}_0$ stands for the combination of observed cosmological data from Planck 2018~\cite{Planck:2018vyg}, DESI DR2~\cite{DESI:2025zgx} and Pantheon+ sample~\cite{Brout:2022vxf}.
Finally, we find that, except for the unconstrained $w_5$ and $w_6$, our reconstruction of $w(z)$ marginally favors dynamical DE in the first bin and is consistent with the cosmological constant at $68\%$ C.L. in the other bins.

Here we reconstruct $w(z)$ with $7$ bins, but $w_5$ and $w_6$ in the last two bins cannot be well constrained. It should be noted that only the $7$ bins and the unconstrained $w_5$ and $w_6$ are not the limitations of our reconstruction method, but a result of the sparse BAO and SNIa data at $z\gtrsim 1$. In other words, if there are more dense BAO and SNIa data at $1\lesssim z\lesssim 2$, our reconstruction can contain more bins at the same time, but the computational cost would not increase much. This is not the case for traditional explicit likelihood inference by globally fitting CMB, BAO and SNIa data because it scales poorly with the dimensionality of the parameter space.

\vspace{5mm}
\noindent {\bf Acknowledgments}
We acknowledge the use of HPC Cluster of Tianhe II in National Supercomputing Center in Guangzhou. The research work is supported by the National Natural Science Foundation of China (12175095), and supported by LiaoNing Revitalization Talents Program (XLYC2007047) .

%%%%%%%%%%%%%%%%%%%%%%%%%%%%%%%%%%%%%%%%
%%%%%%%%%%%%%%%%%%%%%%%%%%%%%%%%%%%%%%%%

%%%%%%%%%%%%%%%%%%%%%%%%%%%%%%%%%%%%%%%%
%%%%%%%%%%%%%%%%%%%%%%%%%%%%%%%%%%%%%%%%

%%%%%%%%%%%%%%%%%%%%%%%%%%%%%%%%%%%%%%%%
%%%%%%%%%%%%%%%%%%%%%%%%%%%%%%%%%%%%%%%%
\end{document}